\documentclass[iop,apj,tighten]{emulateapj}
\usepackage{epstopdf}
\usepackage{amsmath,amstext}

\usepackage{multirow}
\usepackage{amsmath}

\slugcomment{}

\shorttitle{Outflow in suppercritical accretion flows}
\shortauthors{F. Z. Zeraatgari, S. Abbassi \and A. Mosallanezhad}

\begin{document}
\title
{The influence of outflow in supercritical accretion flows}

\author {Fatemeh Zahra Zeraatgari\altaffilmark{1}, Shahram Abbassi\altaffilmark{1,2} and Amin Mosallanezhad\altaffilmark{3,4}}
\affil{$^1$ Department of Physics, School of Sciences, Ferdowsi University of Mashhad, Mashhad, 91775-1436, Iran; fzeraatgari@yahoo.com, abbassi@um.ac.ir}
\affil{$^2$ School of Astronomy, Institute for Research in Fundamental Sciences (IPM), Tehran, 19395-5531, Iran}
\affil{$^3$ Shanghai Astronomical Observatory, Chinese Academy of Sciences, 80 Nandan Road, Shanghai 200030, China; amin@shao.ac.cn}
\affil{$^4$ University of Chinese Academy of Sciences, 19A Yuquan Road, Beijing 100049, China}

\begin{abstract}
We solve the radiation-hydrodynamic (RHD) equations of supercritical accretion flows in the presence of radiation force and outflow by using self similar solutions.  Compare with the pioneer works, in this paper we consider power-law function for mass inflow rate as $ \dot{M} \propto  r^{s} $. We found that $ s = 1 $ when the radiative cooling term is included in the energy equation. Correspondingly,  the effective temperature profile with respect to the radius was obtained as $ T_{\text{eff}} \varpropto r^{-1/2} $.  In addition, we investigated the influence of the outflow on the dynamics of the accretion flow. We also calculated the continuum spectrum emitted from the disk surface as well as the bolometric luminosity of the accretion flow. Furthermore, our results show that the advection parameter, $ f $, strongly depends on mass inflow rate.
\end{abstract}


\section{INTRODUCTION}

It is now widely believed that mass accretion onto black hole is a fundamental and fascinating process for understanding active phenomena in the universe such as active galactic nuclei (AGNs), X-ray binaries (XRBs), gamma-ray bursts (GRBs) and so on.  According to the temperature, black hole accretion disks can be divided into two distinguish branches, i.e., cold and hot disks. For instance, the standard thin disk model (Shakura \& Sunyaev 1973) is a geometrically thin and optically thick accretion disk and belongs to the cold disk group. In fact, the basic idea of this model is that the heat generated via viscosity can be radiated away locally and the accretion disk becomes cold efficiently (the flow temperature is far below virial temperature). Moreover, the standard disk model emits multi-temperature blackbody emission. Additionally, the mass accretion rate of standard disk is mildly low, i.e., $ \dot{M} \lesssim \dot{M}_\text{crit}(\equiv \eta \dot{M}_\text{Edd}) $, where $ \dot{M}_\text{crit} $ denotes critical mass accretion rate, and $ \dot{M}_\text{Edd} $ is the Eddington accretion rate with $ \eta \thicksim 0.1 $ as the radiative efficiency. The usual luminous AGNs and the high/soft state of black-hole binaries belong to this branch (see, reviews by Pringle 1981; Frank et al. 2002; Kato et al. 2008; Abramowicz \& Fragile 2013; Blaes 2013; Lasota 2015 for more details).

In both low-luminosity and also in high-luminosity regimes, the standard-disk picture breaks down. When the mass accretion rate is very low ($  \dot{M} < (0.1-0.3)\, \alpha^{2} \dot{M}_\text{Edd} $),  where $ \alpha $ is the viscous parameter, only a small fraction of dissipated energy radiates away locally. Therefore, due to the inefficient cooling,  most of the generated heat is stored in the accreting gas and advected to the central black hole. Then, the flow temperature becomes extremely high. This new regime is named optically thin Advection-dominated accretion flows (ADAFs). These radiatively inefficient systems can be applied to the low luminosity AGNs and black hole X-ray binaries in hard state (see e.g., Narayan \& Yi (1994, 1995a, 1995b); Abramowicz et al. 1995; Yuan \& Narayan 2014 for more details).

In contrast, when the mass accretion rate is extremely high (approaches or exceeds critical accretion rate), i.e., $ \dot{M} \gtrsim \dot{M}_\text{crit} $, accreting flow becomes very thick in the optical manner and cannot radiate the energy released locally. The radiation then is trapped and advected inwardly with the accreting gas. Such a high-$\dot{M} $ accretion flow called \emph{supercritical} accretion flow or slim disk model (Abramowicz et al. 1988). Indeed, supercritical accretion flows belong to the class of cold disks and similar to the standard disk emit blackbody-like emission. Although, in the presence of photon trapping, which it reduces the radiative efficiency, they are different from standard thin disk model (see e.g., Kats 1977). As a matter of fact, photon trapping takes place when the photon diffusion time in the vertical direction (the time interval of traveling photons from the equatorial plane to the disk surface) exceeds the accretion timescale in radial direction. Therefore, photons are not able to escape from the surface of the disk and then together with gas flow advect towards the central black hole (see, Watarai 1999, 2006; Ohsuga 2002; Ohsuga 2005; Kato et al. 2008 for more details). These optically thick systems may be applied to Ultra-luminous X-ray sources (ULX), microquasars, luminous quasars with luminosity greater than Eddington luminosity, super-soft X-ray sources, and narrow-line seyfert 1 galaxies (Fukue 2004; Kato et al. 2008).

Many analytical works in one/two dimension(s) have been done on supercritical accretion flows to explain the main properties of such systems, e.g., radial velocity, angular velocity, temperature and etc. (e.g., Begelman \& Meier 1982; Abamowicz et al. 1988;  Watarai \& Fukue 1999; Watarai et al. 2000, 2001, 2006; Wang \& Zhou 1999; Mineshige et al. 2000; Fukue 2004; Gu \& Lu 2007; Gu 2012).  In most of the above mentioned theoretical works, due to the technical difficulties, the mass accretion rate of the flow is assumed to be independent of radius. This assumption means that, all the available gas at the outer boundary of the disk can fall down to the black hole horizon and will not escape in the form of wind or jet. However, observations show existence of outflow in black hole accretion flows (e.g., Quataert \& Gruzinov 2000; Bower et al. 2003; Crenshaw et al. 2003; Marrone et al. 2007; Tombesi et al. 2010, 2011a, 2012). This has led to the study of accretion with outflow. In this regards, a substantial number of researchers have investigated the influence of the outflows in accretion disks in simulation aspects (e.g., Stone et al. 1999; Ohsuga et al. 2005, 2009; Ohsuga \& Mineshige 2007, 2011; Hirose et al. 2009; Yuan et al. 2012a,b; Jiang et al. 2013; Yang et al. 2014) and also theoretical works (e.g., Blandford \& Begelman 1999, 2004; Fukue 2004; Mosallanezhad et al. 2013, 2014, 2016; Gu 2015; Samadi et al. 2014, 2016). According to aforementioned works, outflow is a plausible mechanism to carry off mass, angular momentum, and energy from the disk. Then, the dynamics and structure of the accretion flow will be changed significantly in the presence of the wind.

Three possible mechanisms have been introduced to explain the origin of the wind from accretion disks. (1)  magnetocentrifugal outflow where the magnetic fields threading the accretion disk accelerate gas particles (e.g., Blandford \& Payne 1982; Emmering et al. 1992; Miller et al. 2006; Yuan et al. 2015 ). (2) radiation-driven outflow is acted on electrons and lines (see, e. g., Icke 1980; Shlosman \& Vitello 1993; Fukue 2004; Proga 2002), and (3) thermally driven outflow when the thermal velocity of the gas becomes greater than the escape velocity (Begelman et al. 1983; Woods et al. 1996; Sim et al. 2010; Higginbottom \& Proga 2015).

Among a large number of theoretical works done on accretion disk models with wind, Begelman \& Blandford (1999; 2004) presented a creative global analytical solution named adiabatic inflow-outflow solution (ADIOS).  In their solutions they considered radial dependency for mass inflow rate as $ \dot{M} \propto  r^{s} $, with $  0 \leqslant s < 1 $. Recently, in the case of radiatively inefficient accretion flow, Begelman 2012 reformulated his previous works and found that $ s = 1 $. This is also partially supported by recent numerical simulations on optically thin ADAFs (Yuan 2012a,b). In contrast to the optically thin ADAFs, in terms of supercritical accretion disks, the radiative force and perhaps the radiative cooling are not negligible and radiative hydrodynamic equations (RHD) are required to be solved. Therefore, considering power-law function mass inflow rate as $ \dot{M} \propto  r^{s} $ and solving the inflow-outflow equations in the case of supercritical accretion disks might be a valuable work.

The main aim of our present work is to study supercritical accretion flows in the presence of the outflow. As we mentioned above, we adopt power-law function for mass inflow rate. In this work,  the 1.5 dimensional inflow-outflow equations of supercritical  accretion flows in the presence of wind and radiation force will be solved.  Furthermore, to show the angular momentum and energy transfer via outflow we follow the method described in Xie \& Yuan 2008 and Bu et al. 2009.
Then, the flow equations are integrated vertically in cylindrical coordinates. Additionally, to examine the dynamics and structure of the disk, self-similar formalism is assumed.

The outline of this paper is as follows. In the next section, we introduce the basic equations. The self-similar solutions are given in section 3. In section 4, the numerical results are presented and explained in details. Finally, a brief summary and conclusions are provided in section 5.

\section{BASIC EQUATIONS}

In the present work, we investigate structure and dynamics of supercritical accretion flows in which outflows play an important role and carry off mass, angular momentum, and energy from the disk. Following Xie \& Yuan 2008 and Bu et al. 2009 methodology, by adopting self similar approach, we describe the 1.5 dimensional inflow-outflow equations. The mass, momentum and energy conservation equations are integrated vertically in cylindrical coordinates $ (r,\phi , z) $. The optically thick flow is assumed to be axisymmetric, $ (\partial /\partial \phi =0) $, and in a steady state, $ (\partial /\partial t = 0) $. For simplicity, the Newtonian potential, $ \psi (r) = - (GM)/r $, is considered which is more convenient for the self-similar formalism. Further, we neglect the relativistic effects and also self gravity of the accreting gas surrounded the central back hole.  Therefore, the equations of conservation of mass, radial momentum, and angular momentum will be written as follows,

\begin{equation}\label{continuity}
	\dot{M}(r) =  - 2\pi r v_{r}\Sigma,
\end{equation}

\begin{multline}\label{radial momentum}
	v_{r}\frac{dv_{r}}{dr}+\frac{1}{2\pi r\Sigma}\frac{d\dot{M}(r)}{dr}\left(w_{r} - v_{r}\right) = \\
\frac{v_{\phi}^2}{r}-\frac{GM}{r^2}-\frac{1}{\Sigma}\frac{d\Pi}{dr},
\end{multline}

\begin{multline}\label{angular momentum}
  \frac{\Sigma v_{r}}{r}\frac{d}{dr}\left(rv_{\phi}\right)+\frac{1}{2\pi r}\frac{d\dot{M}(r)}{dr}\left(w_{\phi} - v_{\phi}\right) = \\
\frac{1}{r^2}\frac{d}{dr}\left(r^2T_{r\phi}\right),
\end{multline}
where  $ v_{r} $ and $ v_{\phi} $ are radial and rotational velocities, $ \Sigma $ is the vertically integrated density $ (\Sigma \equiv \int \rho\, d\text{z)} $ and $ \Pi $ is the vertically integrated total pressure ($ \Pi  \equiv \int p\, d \text{z} $). In the continuity equation, Equation (\ref{continuity}), $ \dot{M}(r) $ denotes mass inflow rate which is not radially constant and varies with radius. The second terms in the left hand side of Equations (\ref{radial momentum}) and (\ref{angular momentum}) represent  momentum transport outward via outflow. To parameterize the effects of the outflow in angular momentum transport, $ w_{r}=\xi_{1} v_{r} $ and $ w_{\phi}=\xi_{2}v_{\phi} $ are also defined. Noting that, we assume the accreting matter only moves toward the central black hole radially while for the outflow, $ r $ and $ z $  components of velocity are admitted (see figure (1) and appendix of Xie \& Yuan 2008 for more details). In equation (\ref{angular momentum}), only $ r\phi $ component of viscosity is considered, i.e., $ T_{r \phi} = - \alpha \Pi $.  It should be emphasized that in a real case, the magnetic stress driven by the magneto-rotational instability (MRI) transfers the angular momentum outside the disk (Balbas \& Hawley 1991, 1998). Since in our radiation-hydrodynamic (RHD) case,  the magnetic field in not included, the anomalous shear stress tensor has been considered to mimic the magnetic stress.

The hydrostatic balance in the vertical direction is expressed as,

\begin{equation}\label{hydrostatic balance}
  \frac{GM}{r^3}H^2 = \frac{\Pi}{\Sigma} = c_{s}^2,
\end{equation}
where $H$ is the disk half-thickness and $ c_{s} $ being isothermal sound speed.

As it is mentioned in the introduction, in the most of the previous analytical works on the case of supercritical accretion flows, mass inflow rate is assumed to be radially constant. This assumption means that all the available gas at the outer boundary of the disk can fall down to the black hole horizon. On the other hand, both observations and numerical simulations show the existence of outflows in the radiation inefficient accretion flows (RIAFs). For example in terms of  supercritical accretion flows simulations preformed by Ohsuga et al. 2005, 2009, Ohsuga \& Minishige 2011 and Yang et al. 2014 clearly showed that the matter leaves the accretion disk in the form of wind mainly due to the strong radiation pressure force. Therefore, when the outflow is considered, the mass accretion rate will not be constant radially and decrease with decreasing radius. To formulate the effects of wind in our theoretical study, following pioneer works  done by Blandford \& Begelman 1999, 2004, Yuan et al. 2012a,b we consider the mass inflow rate decreases with decreasing radius as,

\begin{equation}\label{mass inflow rate}
	\dot{M}(r) = \dot{M}(r_\text{out}) \left(\frac{r}{r_{\text{out}}} \right)^{s},
\end{equation}
where $ \dot{M}(r_\text{out}) $ is the mass inflow rate at the outer boundary, $ r_\text{out} $.  We also define the dimensionless mass inflow rate at the outer boundary as $   \dot{m}  =  \dot{M}(r_\text{out}) / \dot{M}_\text{crit} $. Here, the critical mass accretion rate is expressed as,

\begin{equation}\label{critical mass rate}
	\dot{M}_\text{crit} =  \frac{L_\text{E}}{c^{2}} = \frac{4 \pi GM}{ c \kappa_{\text{es}}},
\end{equation}
where, $ L_\text{E} $ is the Eddington luminosity, $ \kappa_\text{es}(=\sigma_{T} /m_{H}) $ is the electron scattering opacity, and $ c $ is the speed of light. For the energy equation, we solve the full energy equation. Then, the vertical integration of the energy equation becomes,

\begin{equation}\label{energy}
  Q_\text{adv}^{-} = Q_\text{vis}^{+} - Q_\text{rad}^{-},
\end{equation}
where $ Q_\text{adv}^{-} $ is the advective cooling written as follows,
\begin{multline}\label{advecting cooling}
Q_\text{adv}^{-} =\frac{\Sigma v_{r}}{\gamma -1}\frac{dc_{s}^2}{dr}-2Hc_{s}^2v_{r}\frac{d\rho}{dr} + \\ \frac{1}{2\pi r}\frac{d\dot{M}(r)}{dr}\left(\epsilon_{w}-\epsilon\right).
\end{multline}

Here, $ \epsilon_{w}(=\xi_{3}\epsilon ) $ and $ \epsilon $ are the outflow and inflow specific internal energy, respectively (see, Xie \& Yuan 2008, Bu et al. 2009 for more details). $ \gamma $ denotes the ratio of specific heats. Last term in Equation (\ref{advecting cooling}) shows energy loss via outflow.  In the optically thick regime, the vertical integration of the total pressure $ \Pi $ is sum of the radiation pressure $ \Pi_\text{rad} $ and the gas pressure $ \Pi_\text{gas} $ as,

\begin{equation}\label{equation of state}
 \Pi = \Pi_\text{rad} + \Pi_\text{gas}.
\end{equation}

In this work, we focus on the regime where the radiation pressure is much higher than the gas pressure (radiation pressure-supported disk), and therefore we neglect the gas pressure throughout this paper. So, this leads to $ \gamma = 4/3 $. In the equation (\ref{energy}), $ Q_\text{vis}^{+} $ is viscous heating rate generated via viscosity and we adopt the following form,
\begin{equation}\label{viscous heating}
Q_\text{vis}^{+} = r T_{r\phi} \frac{d\Omega}{dr},
 \end{equation}
where  $ \Omega(\equiv v_{\phi}/r) $ represents the angular velocity of disk rotation. To formulate the radiative cooling rate, $ Q_\text{rad}^{-} $, we first assume, in the optically thick disk, the heat loss in the vertical direction is much higher than exchanged heat in the radial direction. Secondly, since the disk is considered to be radiation-supported disk, for simplicity, we consider only electron scattering for the opacity $ \kappa_{\text{es}} $ and neglect the free-free opacity, $ \kappa_{\text{ff}} $, i.e.,  $ \bar{\kappa} \simeq \kappa_{\text{es}}  $  . Then, the radiative cooling rate will be approximately written as,

\begin{equation}\label{radiative cooling}
Q_\text{rad}^{-} =\frac{8acT_{0}^4}{3 \bar{\kappa} \rho H}\simeq \frac{8\, c\, \Pi}{\kappa_{\text{es}} \Sigma H},
\end{equation}
where $ a $ is the radiation constant and $  T_{0} $ represents the disk temperature on the equatorial plane.

\section{Self-Similar Solutions}

We assume that the physical quantities are self-similar in the radial direction. In the self-similar formalism the velocities can be expressed as follows

\begin{equation}\label{self-similar vr}
  v_{r} = - c_{1} v_\text{ko}   \left(\frac{r}{r_\text{out}} \right)^{-1/2},
\end{equation}

\begin{equation} \label{self-similar vp}
  v_{\phi} = c_{2} v_\text{ko}\left( \frac{r}{r_\text{out}} \right)^{-1/2},
\end{equation}

\begin{equation}\label{Keprerian velocity}
  v_\text{ko}=\sqrt{\frac{GM}{r_\text{out}}},
\end{equation}
where, $ v_\text{ko} $ represents the Keplerian velocity at the outer boundary, $ c_{1} $ and $ c_{2} $ are constants which will be determined later. By substituting Equations (\ref{mass inflow rate}), (\ref{self-similar vr})-(\ref{Keprerian velocity}) into equations (\ref{continuity}) and (\ref{angular momentum}) we can simply find an explicit expression for the vertically integrated pressure as,

\begin{equation}\label{pressure}
  \Pi = \frac{\dot{M} \Omega}{2 \pi \alpha} \left[ \frac{1 - 2s (\xi_{2}-1)}{2s + 1} \right].
\end{equation}

The above relation shows the influence of the outflow in pressure equation which in our case this pressure denotes the radiation pressure. With considering constant mass accretion rate, i. e., $ s = 0 $, this expression is reduced to equation (9.1) of Kato (2008). In addition, by adopting the vertically integrated density from continuity equation and using pressure relation, Equation \ref{pressure}, we can easily obtain scale-height of the disk from hydrostatic balance equation (Equation \ref{hydrostatic balance}) as,

\begin{equation}\label{scale height}
 \left( \frac{H}{r} \right)^{2} = \frac{c_{1} c_{2}}{\alpha} \left[  \frac{1 - 2s (\xi_{2}-1)}{2s + 1}  \right].
\end{equation}

Now we have radially self-similar equations for $ H $, $ \Sigma $, and $ \Pi $  which only depend on $ c_{1} $ and $ c_{2} $. We thus need a system of two equations to obtain $ c_{1} $ and $ c_{2} $. Hence, by substituting our self similar solutions into the energy equation, very surprisingly, it is found that the cooling and heating rates have the same radial dependency only if $ s = 1 $. This is totally in agreement with Begelman (2012)\footnote{It should be noted here that in the recent paper of Begelman (2012) he found, in terms of adiabatic inflow-outflow solution (ADIOS) model for radiatively inefficient accretion flows (RIAF), the mass flux satisfies $ \dot{M} \propto r^{n} $ with $ n = 1 $. Here we want to show that this prediction is not only true in the case of ADAFs but also is satisfied in terms of supercritical accretion disk where radiation pressure play an important role.}. Thus, for the rest of our calculations we set $ s = 1 $. The energy equation is reduced to,

\begin{multline}\label{self-similar energy}
	\left[ \left(6 \xi_{3} - 1 \right) c_{1} -3\alpha c_{2} \right]   \sqrt{3\left( 3 - 2 \xi_{2} \right) c_{2}}\, \dot{m}\, + \\
	48 \sqrt{\alpha c_{1}} \left( \frac{r_\text{out}}{r_{s}} \right)  = 0.
\end{multline}

Also, in similar way, the following relation can be obtained from the radial momentum conservation,

\begin{equation}\label{self-similar radial momentum}
3\alpha \left(2 \xi_{1} - 1\right) c_{1}^2 + \left( 3 - 2 \xi_{2} \right) c_{1}c_{2} + 6\alpha \left( c_{2}^2 - 1 \right) = 0.
\end{equation}

For given values of $ \alpha $ , $ \xi_{1} $, $ \xi_{2} $, $ \xi_{3} $, $ \dot{m} $  and $ r_\text{out} $, the set of equations (\ref{self-similar energy}) and (\ref{self-similar radial momentum}) can be solved to determine the dynamics of the accretion flow.

\section{numerical results}

\begin{figure*}
\begin{center}
\includegraphics[width=\textwidth, angle=0]{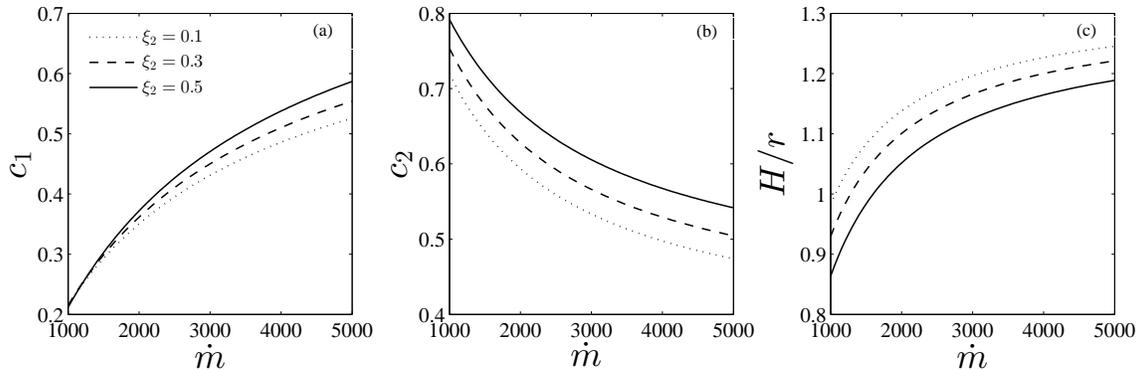}
\caption{The dynamics of the accretion flow with mass inflow rate at the outer boundary $ \dot{m} $ for $ \xi_{2} =  $ 0.1 (dotted line), 0.3 (dashed line), and 0.5 (solid line) in radius $ r_\text{out} = 100 r_{s} $. Other parameters are set to be $\xi_{1} = 0.5 $, $ \xi_{3} = 0.2 $ and $ \alpha = 0.1 $.}
\label{variables_xi2}
\end{center}
\end{figure*}

\subsection{Dynamical solutions}

In our calculations we set $ \alpha = 0.1 $, $ r_\text{out} = 100 r_{s} $, $ M = 10^6 M_{\odot} $, where $ M_{\odot}  $ is solar mass. We solved equations (\ref{self-similar energy}) and (\ref{self-similar radial momentum}) to obtain the behavior of the physical quantities which are shown in Figure \ref{variables_xi2}. In three panels of Figure \ref{variables_xi2} the variations of radial velocity (a), rotational velocity (b), and half-thickness of the accretion disk (c) with respect to the mass inflow rate at the outer boundary, $ \dot{m} $, for different values of angular momentum of the outflow, $ \xi_{2} =  $ 0.1 (dotted line), 0.3 (dashed line), and 0.5 (solid line), have been plotted. As it can be seen, radial velocity and half-thickness of the accretion disk increase with increasing of mass inflow rate at the outer boundary, $ \dot{m} $, while rotational velocity decreases. Additionally, radial and angular velocities of the disk have an increasing trend with respect to the angular momentum of the outflow, $  \xi_{2} $, and both are sub-Keplerian. This is because, outflow can take away more angular momentum from the disk and more matter can inflow to the black-hole. Consequently, the radial velocity increases with $ \xi_{2} $.  Also, The half-thickness of the disk decreases when the $ \xi_{2} $ increases.  
\subsection{Radiation Properties and Continuum Spectrum}

It is assumed that the surface of the disk radiates locally the blackbody radiation,

\begin{equation}\label{black body}
	B_{\nu}(r) = \frac{2h}{c^{2}} \frac{\nu^3}{\text{e}^{ h\nu/k_\text{B}T_{\text{eff}}(r) } -1}
\end{equation}
which is multi-color (different temperatures at different radii).

The emergent local flux from the surface of the disk is

\begin{equation}\label{flux}
	F = \frac{1}{2} Q_\text{rad}^{-} = \frac{16 \sigma T_{0}^{4}}{3 \tau} = \sigma T_{\text{eff}}^{4}\, ,
\end{equation}
where $ \tau\, (= \kappa_\text{es} \Sigma/2) $ is the optical depth, $ \sigma $ is the Stefan-Boltzmann constant, and the factor 2 represents radiation from the two sides of the disk. Therefore, the effective temperature of the disk surface is obtained as,

\begin{equation}\label{temperature}
	\sigma T_{\text{eff}}^4 =\frac{4c}{\kappa_\text{es}}\frac{GM}{r^2}\sqrt{\frac{\left( 3 - 2\xi_{2} \right) c_{1} c_{2} }{3\alpha}}\,,
\end{equation}

\begin{figure}
\begin{center}
\includegraphics[width=90mm]{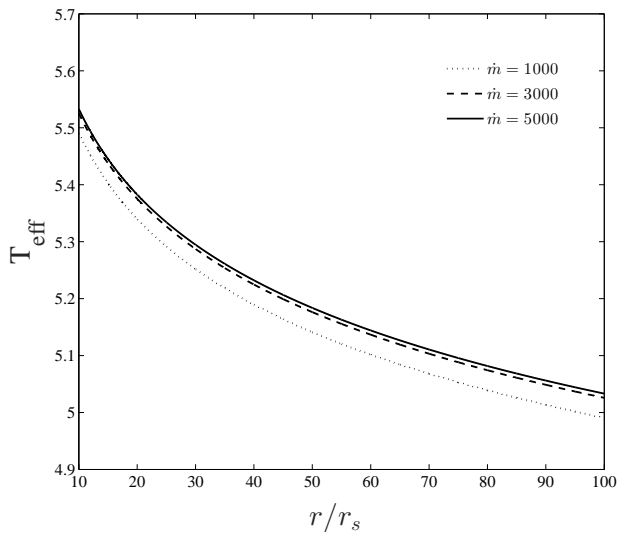}
\caption{Variation of the effective temperature with radius for $ \dot{m} $ = 1000 (dotted line), 3000 (dashed line), and 5000 (solid line). Here, $ \xi_{1} = \xi_{2} = 0.5 $, $ \xi_{3} = 0.2 $, and $ \alpha = 0.1 $.}
\label{temp_radius}
\end{center}
\end{figure}

It is clear that the temperature is proportional to $ r^{-1/2} $, a flatter temperature profile compared to the standard disk model, which is totally in agreement with the solutions for the critical accretion disks model represented by Fukue (2004). Additionally, it should be noted here that the effective temperature depends on the accretion rate as shown in Figure (\ref{temp_radius}) for three values of mass inflow rate at the outer boundary, $ \dot{m} = 1000$ (dotted line), $3000 $ (dashed line), and $5000$ (solid line). As you can see the effective temperature increases as the mass inflow rate at the outer boundary increases, while it decreases over the full range of radii.

The continuum spectrum (luminosity per frequency) of the disk surface is calculated by integrating the blackbody radiation over the surface of the disk in the range of $ r_\text{in} \leqslant r \leqslant r_\text{out} $ as,

\begin{equation}\label{continuum spectrum}
	L_{\nu} = 2 \int_{r_\text{in}}^{r_\text{out}} \pi B_{\nu}(r)  2\pi\, r\, dr ,
\end{equation}
where a factor 2 means both sides of the disk.

\begin{figure}
\begin{center}
\includegraphics[width=90mm]{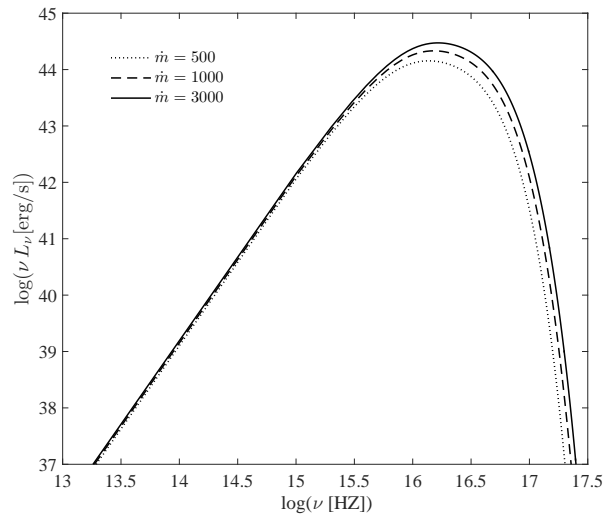}
\caption{Continuum spectra of suppercritical accretion disks. The central black-hole mass is fixed as $ 10^6 M_{\odot} $, while the mass accretion rate at the outer boundary is $\dot{m} = $ 500 (dotted line), 1000 (dashed line), and 3000 (solid line).}
\label{lum_nu}
\end{center}
\end{figure}

The continuum spectra of the supercritical accretion flow has been also plotted in figure \ref{lum_nu} for a typical central black hole with mass $ M =10^6M_{\odot} $, $ r_\text{in}= 10r_{s} $, $ r_\text{out}=100r_{s} $, and various mass inflow rates at the outer boundary, $ \dot{m} = $ 500 (dotted line), 1000 (dashed line), and 3000 (solid line). It should be noted that the maximum of the $ \nu L_{\nu} $ is almost of the order Eddington luminosity of the central black hole.

\subsection{The Bolometric Luminosity}

The bolometric luminosity, $ L $, can be calculated for supercritical accretion disks using one dimensional solutions represented in this work. Then, this quantity can be written as,

\begin{equation}\label{bolometric luminosity}
  L = 2\int_{r_\text{in}}^{r_\text{out}} F(r)\, 2 \pi r dr.
\end{equation}

\begin{figure}
\begin{center}
\includegraphics[width=90mm]{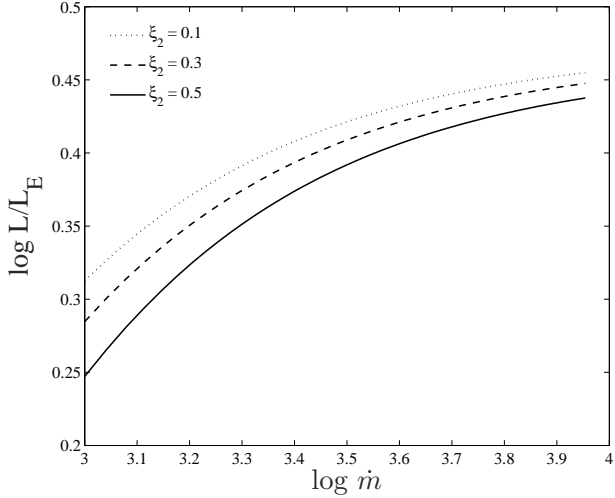}
\caption{Bolometric luminosity as a function of mass inflow rate at the outer boundary for $ \xi_{2} = 0.1 $ (dotted line), $ 0.3 $ (dashed line), and $ 0.5 $ (solid line). Here, $ \xi_{1} = 0.5 $, $ \xi_{3} = 0.2 $,  and $ \alpha = 0.1 $.}
\label{lum_mdot}
\end{center}
\end{figure}

The resulting profile as a function of mass accretion rate is plotted in figure \ref{lum_mdot} for different amounts of angular momentum carried by the outflow, $ \xi_{2} = 0.1$ (dotted line), $ 0.3 $ (dashed line), and $ 0.5 $ (solid line). It is clear that when the amount of angular momentum carried by the outflow increases the luminosity of the supercritical accretion disk decreases and becomes somewhat around the Eddington luminosity of the central black hole. In addition, the disk luminosity is almost insensitive to mass inflow rate at the outer boundary and is nearly close to $ L_\text{Edd} $. Moreover, observed luminosity depends on viewing angle because the emission is mainly toward the vertical direction of the accretion disk. It should be emphasized that we can observe a super-Eddington luminosity when the viewing angle of the disk is face-on.  On the other hand, sub-Eddington luminosity will be measured in edge-on viewing angle (Watarai 2006).

\subsection{Radiative-Cooling-dominated Regime}
We are going to examine the influence of the outflow on the advection parameter, $ f $, definded by Narayan \& Yi 1994, which is the ratio of the advective cooling to the viscous heating. In our solution, $ f $ parameter can be obtained as,

\begin{equation}\label{advection parameter}
  f \equiv \frac{Q_\text{adv}^{-}}{Q_\text{vis}^{+}} = \frac{\left( 6\xi_{3} - 1 \right) c_{1}}{3\alpha c_{2}}\, .
\end{equation}

\begin{figure}
\begin{center}
\includegraphics[width=90mm]{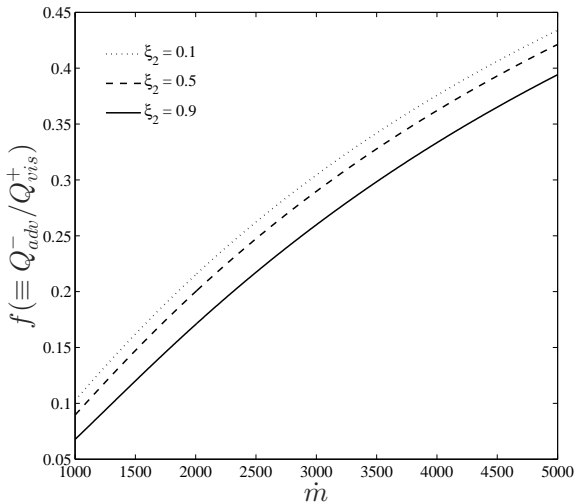}
\caption{Variation of advection parameter, $ f $, as a function of mass inflow rate at the outer boundary. $\xi_{2} =$ 0.1 (dotted line), 0.5 (dashed line), and 0.9 (solid line). }
\label{f_adv}
\end{center}
\end{figure}

The result has been shown in figure \ref{f_adv} with respect to the $ \dot{m} $ for various angular momentums contributed to the outflow, $\xi_{2} =$ 0.1 (dotted line), 0.5 (dashed line), and 0.9 (solid line). As it is shown, advection parameter $ f $ increases as $ \dot{m} $ increases at the outer boundary of the disk. Moreover, $ f $ and $ \xi_{2} $ vary inversely with each other, i.e., angular momentum transport leads to reduction of advection in radial direction\footnote{It should be noted here that Zeraatgari \& Abbassi 2015 also investigated the effect of advection parameter along the vertical direction. They found that the advection parameter is not fixed along $ \theta $ and reaches to its maximum near the rotation axis.}.This can be simply understand from Equation (\ref{advection parameter}).


\section{Brief Summery and conclusions}

In this paper, we have studied the influence of the outflow on the dynamics and the structure of supercritical accretion flows. Following Xie \& Yuan 2008 and Bu et al. 2009 methodology, we described the 1.5 dimensional inflow-outflow equations. For simplicity, the central black hole gravity was described as the Newtonian potential. The radiation-hydrodynamic equations (RHD) were considered and for the radiative cooling only electron-scattering diffusion in the vertical direction was assumed. Then, the mass, momentum, and energy conservation equations were integrated vertically in cylindrical coordinates. We considered power-law function for mass inflow rate as $ \dot{M} \propto  r^{s} $, and solved the inflow-outflow equations by using self similar approach in the supercritical accretion regime. Additionally, we posited some mass, angular momentum, and energy contributed to the outflow. Therefore, we examined how the outflow affects on the dynamics of the supercritical accretion disk. Consequently, we found that with increasing the angular momentum of the outflow the radial and rotational velocities of the disk increase while the height-thickness of the accretion disk decreases. The effective temperature represented in this solution is a function of radius as $ T_{\text{eff}}\varpropto r^{-1/2} $ which is reduced from inner regions of the disk to outer regions. In addition, the effective temperature has a rising behavior in respect to the mass inflow rate at the outer boundary. On the other hand, increasing mass inflow rate makes the temperature of the disk increase gradually. We also calculated the continuum spectrum emitted from surface of the disk with assuming blackbody radiation . The isotropic luminosity decreases as $ \dot{m} $ at the outer boundary decreases and becomes softer. The variation of bolometric luminosity against $ \dot{m} $ shows the luminosity is kept around the Eddington luminosity for high $ \dot{m} $ that is in agreement with simulations done by Ohsuga (2005). The self-similar solutions represented in this paper indicate that $ f $ strongly depends on mass inflow rate and an increase in $ \dot{m} $ can dramatically magnify the advection radially. 

\subsection{Future Work}
In the radiation-hydrodynamic equations of supercritical accretion disk presented here, we have used radially one-dimentinal self-similar solutions. In spite of simplifications considered in this model, our results give us a better understanding of such a complicated system. However, for more accurate examination of the outflow, it is significantly superior to study such systems in two dimensions. We therefore plan to solve the two-dimensional radiation hydrodynamic equations by using relaxation method (two boundary value problem). For the radiation flux also we will consider full radiative process with radiative transfer along both the radial and vertical direction. Furthermore, following pioneer work (Kate 1977), it is worthy for our future work  to evaluate the fraction of the accretional energy that is swept into the black hole in the form of trapped photons, versus the radiated fraction.

\acknowledgements
The authors thank Feng Yuan for his useful suggestions and discussions. We also appreciate the referee for his/her thoughtful and constructive comments in the early version of the paper. S. Abbassi acknowledges support from the International
Center for Theoretical Physics (ICTP) for a visit through the regular
associateship scheme. This work was supported by Ferdowsi University of Mashhad under grant 3/37875 (1394/04/03).

\end{document}